\documentclass[%
 reprint,
superscriptaddress,
 amsmath,amssymb,
 aps,
]{revtex4-2}

\usepackage{graphicx}
\usepackage{dcolumn}
\usepackage{bm}
\usepackage{xcolor}
\usepackage{braket}
\usepackage{amsmath, amssymb,mathcomp,enumerate}
\usepackage{bbold}
\usepackage{soul}
\usepackage[colorlinks=true,allcolors=black]{hyperref}
\usepackage{comment}
\usepackage{array}

\newcommand{\wa}[1]{\textcolor{teal}{#1}}

\begin{document}

\title{Using bi-fluxon tunneling to protect the Fluxonium qubit
}

\author{Waël Ardati}
\email{wael.ardati@neel.cnrs.fr}
\affiliation{Univ. Grenoble Alpes, CNRS, Grenoble INP, Institut N\'eel, 38000 Grenoble, France}
\author{Sébastien Léger}
\affiliation{Department of Applied Physics, Stanford University, Stanford, California 94305, USA}
\author{Shelender Kumar}
\affiliation{Univ. Grenoble Alpes, CNRS, Grenoble INP, Institut N\'eel, 38000 Grenoble, France}
\author{Vishnu Narayanan Suresh}
\affiliation{Univ. Grenoble Alpes, CNRS, Grenoble INP, Institut N\'eel, 38000 Grenoble, France}
\author{Dorian Nicolas}
\affiliation{Univ. Grenoble Alpes, CNRS, Grenoble INP, Institut N\'eel, 38000 Grenoble, France}
\author{Cyril Mori}
\affiliation{Univ. Grenoble Alpes, CNRS, Grenoble INP, Institut N\'eel, 38000 Grenoble, France}
\author{Francesca D'Esposito}
\affiliation{Univ. Grenoble Alpes, CNRS, Grenoble INP, Institut N\'eel, 38000 Grenoble, France}
\author{Tereza Vakhtel}
\affiliation{Instituut-Lorentz, Universiteit Leiden, P.O. Box 9506, 2300 RA Leiden, The Netherlands}
\author{Olivier Buisson}
\affiliation{Univ. Grenoble Alpes, CNRS, Grenoble INP, Institut N\'eel, 38000 Grenoble, France}
\author{Quentin Ficheux}
\email{quentin.ficheux@neel.cnrs.fr}
\affiliation{Univ. Grenoble Alpes, CNRS, Grenoble INP, Institut N\'eel, 38000 Grenoble, France}
\author{Nicolas Roch}
\email{nicolas.roch@neel.cnrs.fr}
\affiliation{Univ. Grenoble Alpes, CNRS, Grenoble INP, Institut N\'eel, 38000 Grenoble, France}

\date{February 2024}

\begin{abstract}
Encoding quantum information in quantum states with disjoint wave-function support and noise insensitive energies is the key behind the idea of qubit protection. While fully protected qubits are expected to offer exponential protection against both energy relaxation and pure dephasing, simpler circuits may grant partial protection with currently achievable parameters. Here, we study a fluxonium circuit in which the wave-functions are engineered to minimize their overlap while benefiting from a first-order-insensitive flux sweet spot. Taking advantage of a large superinductance ($L\sim 1~\mu \rm{H}$), our circuit incorporates a resonant tunneling mechanism at zero external flux that couples states with the same fluxon parity, thus enabling bifluxon tunneling. The states $\ket{0}$ and $\ket{1}$ are encoded in wave-functions with parities 0 and 1, respectively, ensuring a minimal form of protection against relaxation. Two-tone spectroscopy reveals the energy level structure of the circuit and the presence of $4 \pi$ quantum-phase slips between different potential wells corresponding to $m=\pm 1$ fluxons, which can be precisely described by a simple fluxonium Hamiltonian or by an effective bifluxon Hamiltonian. Despite suboptimal fabrication, the measured relaxation ($T_1 = 177\pm 3 ~\mu s$) and dephasing ($T_2^E = 75\pm 5~\mu \rm{s}$) times not only demonstrate the relevance of our approach but also opens an alternative direction towards quantum computing using partially-protected fluxonium qubits.
\end{abstract}

\maketitle

\section{Introduction}

\begin{figure}

\centering
\includegraphics[width=1\columnwidth]{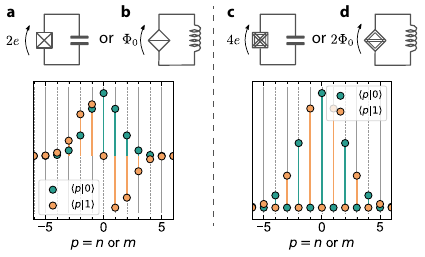}
\caption{\label{concept} \textit{Parity protection in superconducting quantum circuits.} \textbf{a.} Cooper-pair box circuit with ground and first-excited state wave-functions in the integer-charge n basis. \textbf{b.} Flux qubit with a quantum-phase slip element and associated wave-functions in the integer-flux m basis. \textbf{c.} Charge circuit allowing protection against both bit- and phase-flips. The cross-hatched circuit element permits only the tunneling of pairs of Cooper pairs. The associated $\ket{0}$ and $\ket{1}$ wave-functions show no overlap in the charge basis. \textbf{d.} Flux-circuit with a double-quantum-phase slip element. The cross-hatched circuit element permits only the tunneling of pairs of fluxons.
}
\end{figure}

The properties of superconducting quantum circuits are described by two conjugated variables, $\phi$ that generalizes the flux across an inductor and $Q$ the charge accumulated on a capacitance~\cite{Devoret.1995}. The associated energies are the inductive energy $E_\text{L}=(\hbar/2e)^2/L$ and the charging energy $E_\text{C}=e^2/2C$ respectively, with $C$ a capacitance and $L$ an inductance. Quantum information can be encoded into one of these degrees of freedom and the recipe to form a qubit is usually the following: hybridizing charge or flux states that differ by one integer ($N$ and $N+1$ charge or flux states for example). This hybridization is obtained via the coherent tunneling of Cooper pairs or fluxons via an operator, which takes the form $\ket{N}\bra{N+1}+\ket{N+1}\bra{N}$. This is the celebrated Josephson Hamiltonian $-E_\text{J}\cos(\hat \varphi)$ in the case of charge tunneling while its dual, the coherent Quantum Phase Slip (QPS) Hamiltonian $E_\text{S}\cos(2\pi \hat n)$, describes the coherent tunneling between two adjacent flux states \cite{Mooij2005,Mooij2006}. These two configurations are described in figure~\ref{concept}.\textbf{a} and \textbf{b}. The energy spectrum of this charge-like (resp. flux-like) qubit depends on the ratio $E_\text{J}/E_\text{C}$ (resp. $E_\text{S}/E_\text{L}$). The first demonstrated superconducting qubit was operated in the regime $E_\text{C}> E_\text{J}$~\cite{Nakamura.1999}, or \text{Cooper-pair box} regime, which made it very sensitive to charge noise and, as a consequence, very low coherence. In the opposite regime ($E_\text{J}\gg E_\text{C}$), the energy levels of a charge-like qubit show an exponentially small charge dispersion, hence providing an excellent protection against dephasing. It is called the \textit{Transmon} regime~\cite{Koch.2007}. This protection coming from vanishing dispersion can also be obtained for flux-like qubits when $E_\text{S}\gg E_\text{L}$. However, it is harder to reach because of the strong asymmetry between the vacuum permittivity $\epsilon_0$ and permeability $\mu_0$. So far, $E_\text{S} > E_\text{L} \sim 10$ was achieved with the fluxonium in the \textit{quasi-charge} regime~\cite{Pechenezhskiy.2020}. While both qubit types have demonstrated coherence times in the millisecond range~\cite{Pop.2014, Place.2021, Somoroff.2023} and single- and two-qubit gate fidelities above 99.9\%~\cite{Ding.2023,Zhang.2023}, they are not protected against energy relaxation and the road towards an error-corrected quantum processor is still long.

It turns out that the noise-protection against dephasing can be extended to create superconducting qubits that are also robust against energy relaxation. The main idea behind this relaxation protection is to use an extra internal symmetry of the system and encode the ground and excited states with wave-functions of different parities~~\cite{Douçot.2012, Gyenis.2021oc, Douçot.2002, Ioffe.2002o78, Kitaev.2006}. For example in the parity of the number of Cooper pairs or in the parity of the flux quanta. This  parity encoding offers a major advantage: protection against depolarization or energy relaxation. This can be simply understood recalling Fermi Golden Rule, where the relaxation rate $\Gamma_1\equiv 1/T_1$ is proportional to $\left|\bra{g} H_\lambda\ket{e}\right|^2$ with $H_\lambda$ the coupling Hamiltonian to a fluctuating quantity $\lambda$. Since the ground and excited states live in different parity sectors of the qubit phase space, they will have disjoint wave-function supports and this matrix element is suppressed, leading to a potentially infinitely long $T_1$. The first proposals relied on large-scale arrays~\cite{Gladchenko.2008, Bell.2014, Dodge.20235a} and were extremely challenging to design experimentally and even calculating their exact spectrum is a very difficult task~\cite{Weiss.2019}.

In 2013, Brooks, Kitaev and Preskill introduced a circuit with only three independent modes that could offer such full protection: the $0-\pi$ qubit~\cite{Brooks.2013}. Interestingly this circuit can be mapped to an even simpler one: a capacitance in parallel with a circuit element that only allows the tunneling of pairs of Cooper pairs or $\cos(2 \hat{\varphi})$ element~\cite{Paolo.2019mqd} (see Fig.~\ref{concept}.c). As a consequence of this tunneling induced parity, ground and excited states wave-functions of the $\cos(2 \hat{\varphi})$ protected qubit are superpositions of even and odd Cooper pair number states respectively~\cite{Smith.2020}. The dual circuit, also known as the bi-fluxon qubit~\cite{Kalashnikov.2020,Douçot.2012}, can be viewed as a superconducting loop interrupted by a $\cos(4 \pi  \hat{n})$ element allowing only double QPS. In this case the wave-functions are superpositions of even and odd numbers of fluxons (Fig.~\ref{concept}.d). These wave-functions with disjoint supports offer protection against depolarization but, as discussed previously, a full protection also requires suppressed energy dispersion. As with the Transmon and its exponential protection against charge noise conferred by a high $E_\text{J}/E_\text{C}$ ratio, this implies $E_\text{J,2}/E_\text{C}\gg 1$ for the $\cos(2 \hat{\varphi})$ qubit~\cite{Smith.2020} and $E_{S,2}/E_\text{L}\gg 1$ in the case of the $\cos(4 \pi  \hat{n})$ qubit~\cite{Kalashnikov.2020}, where $E_\text{J,2}$ (resp. $E_{S,2}$) is the energy scale associated with tunneling of pairs of Cooper pairs (resp. pairs of fluxons).
 
 This combination of suppressed energy dispersion and disjoint wave-function supports is usually referred to as the \textit{hard} protection regime as opposed to the \textit{soft} case where protection is only partially achieved. To date only the soft regime has been demonstrated experimentally~\cite{Kalashnikov.2020, Gyenis.2021oc} since obtaining a pure $\cos(2 \hat{\varphi})$ or a pure $\cos(4 \pi \hat{n})$ element is extremely challenging. Pushing them to the hard regime requires extremely large inductances or capacitances combined with highly symmetrical electrical circuit elements, which remains extremely challenging for the current technology.

 Here, we show that a fluxonium qubit with a very large superinductance ($L\sim 1~\mu \text{H}$) and biased at zero flux can benefit from bi-fluxon protection, which endows it with disjoint wave-functions while being first order protected against dephasing. Notably the 01 transition of our devices is around $3\ \text{GHz}$ but still we observe $T_1$ lifetime of $177\pm 3~\mu s$ (or equivalently a quality factor $Q\sim 3.4 \times 10^6$) and $T_{2, \text{E}}$ of $75\pm 5~ \mu s$ in a planar geometry without special cleaning or material development. A Transmon with similar dielectric loss and 01 frequency would show an order of magnitude lower coherences. This demonstrates that hardware-protected qubit concepts can be easily implemented in a simple circuit with a direct and significant improvement in coherence.

\section{quantum phase slips in a Fluxonium qubit}

The Fluxonium qubit consists in a Josephson junction with Josephson energy $E_\text{J}$ shunted by a capacitance $C$ and an inductance $L$. Its Hamiltonian reads~\cite{Manucharyan.2009qb1, Koch.2009} :
\begin{equation}
\label{H_fluxo}
\hat H= 4 E_\text{C} \hat n^2 - E_\text{J} \cos \hat\varphi+\frac{E_\text{L}}{2}\left(\hat\varphi+\frac{\phi_\text{ext}}{\varphi_0}\right)^2,
\end{equation}
with $\hat n$ the charge operator in units of Cooper pairs, $\hat\varphi$ the superconducting phase difference across the junction, $E_\text{C}$ the charging energy, $E_\text{L}$ the inductive energy, $\varphi_0=\hbar/2e$ the reduced flux quantum and $\phi_\text{ext}$ the external magnetic flux through the loop. When $E_\text{L}$ is small, the states of the fluxonium can be understood as \textit{plasmons} that are charge oscillations across the Josephson junction and \textit{persistent-current states} or \textit{fluxons}~\cite{Koch.2009}.

\begin{figure*}
\centering
\includegraphics[width=\textwidth]{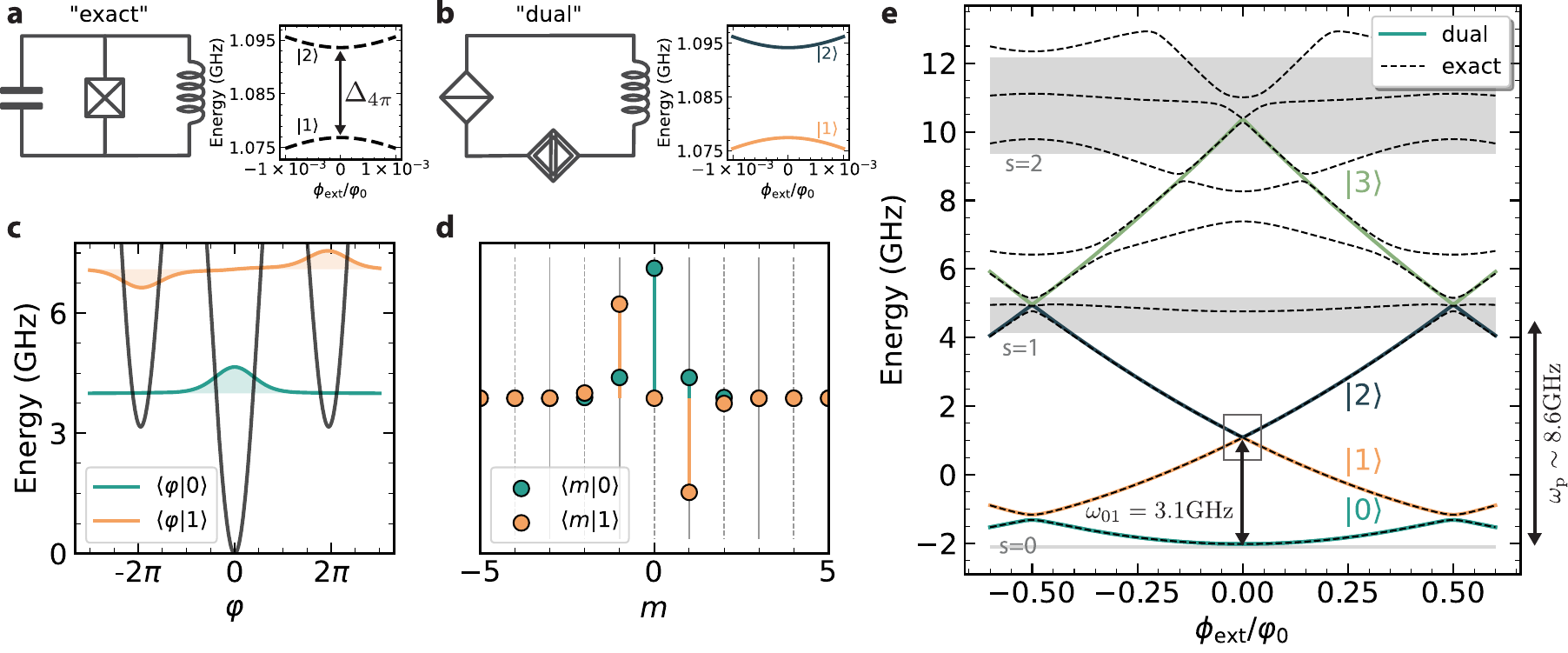}
\caption{\label{qps_&_fluxonium}\textit{Protecting the fluxonium using double Quantum Phase Slips (QPS)}. \textbf{a.} Layout of the fluxonium circuit. At zero-flux and in the $E_\text{L}\ll \sqrt{8E_\text{J}E_\text{C}}$ regime, the first and second excited states are two fluxon states coupled by double QPS. \textbf{b.} Effective circuit representing an inductive loop interrupted by single (diamond) and double (cross-hatched diamond) quantum-phase slip elements. \textbf{c.} Confining potential (black) in the continuous $\varphi$-representation and wave-functions of the ground and excited states of the fluxonium circuit used in this work when biased at $\phi_\text{ext}=0$. \textbf{d.} Identical wave-functions in the discrete fluxon basis. The $\ket{0}$ and $\ket{1}$ states are almost non-overlapping and benefit from energy-relaxation protection thanks to bi-fluxon tunneling. \textbf{e.} Energy spectrum of the circuit obtained form the fluxonium model (Eq.~(\ref{H_fluxo}), dashed line) and the effective "dual" model (Eq.~(\ref{dual}), colored line). At low energy, the parabolic dispersion of the fluxon states is recovered. These states are coupled via QPS induced by the first Bloch-band (grey region, $s=0$), which induces $\Delta_{2 \pi}$ splitting between the ground and excited states at half-integer flux and $\Delta_{4\pi}$ splitting between first and second excited states at zero-flux. At energies higher than the plasma frequency ($\sim \omega_\text{p}=\sqrt{8 E_\text{J} E_\text{C}}$), the dual model cannot predict the correct energy levels. It can be refined to take into account higher Bloch bands and inter-band couplings (see appendix \ref{Bloch}).}
\end{figure*}

Interestingly, if $E_\text{L} \ll \sqrt{8 E_\text{J} E_\text{C}}$ and $E_\text{J}/E_\text{C}\gg 1$, the fluxonium circuit can be effectively described as a superconducting loop of inductance $L$ interrupted by QPS elements~\cite{Mooij.2006} that allow phase winding of multiples of $2\pi$, hence coupling fluxons states (see Fig.~\ref{qps_&_fluxonium}). The Hamiltonian~(\ref{H_fluxo}) can then be approximated to~\cite{ Zhu.20131yu, Ulrich.2016}:
\begin{equation}
    \label{dual}
    \hat H = 2\pi^2 E_\text{L} \left(\hat{m}+\frac{\phi_\text{ext}}{\phi_0} \right)^2
    +E_{S,1} \cos(2\pi \hat{n}  ) + E_{S,2} \cos(4\pi \hat{n}  ).
\end{equation}
Here, $E_\text{L}$ plays the role of the kinetic energy whose associated variable $ \hat m$ is the number of flux quanta trapped in the superconducting loop. The conjugate variable of $\hat m$ is called the \textit{quasi-charge} $\hat n$~\cite{Likharev.1985}. It is periodic on the interval $[0,1]$ and can be used to define the potential energy of the device and corresponds to the first Bloch band (labelled $s=0$ in figure~\ref{qps_&_fluxonium}.\textbf{e}). Here, the potential energy is composed of two terms which describe QPS that change the $m$ value  by $\pm 1$ or $\pm 2$, and  which are respectively represented by a diamond and a cross-hatched diamond in figure~\ref{qps_&_fluxonium}.\textbf{b}. In general the Bloch band is not a pure cosine nor a sum of two cosines~\cite{Likharev.1985, Koch.2009}. Moreover several Bloch bands should be accounted for (they are labelled $s=1$ ans $s=2$ on figure~\ref{qps_&_fluxonium}.\textbf{e}). However for the samples presented in this paper these are good approximations (see appendix~\ref{Bloch}). 

The Fluxonium is usually operated at half-integer external flux~\cite{Nguyen.2019, Grunhaupt.2019, Zhang.2021sf, Bao.2022, Najera-Santos.2023}. At this point, the two lowest-lying states are two fluxons  differing by $\Delta m = 1$ and coupled via single quantum phase slips occurring at the Josephson junction with an amplitude $\Delta_{2\pi}$. The ground and excited states can then be described as bonding and anti-bonding combinations of two persistent-current states of opposite circulating direction. In phase representation, these two wave-functions are odd/even combination of two wave packets centered at $\varphi = \pm \pi$. In this configuration, there is a large overlap of these two wave-functions and the Fluxonium does not benefit from any $T_1$-protection~\cite{Gyenis.2021oc}. At other flux points, it is possible to engineer meta-stable states that show extremely long lifetime~\cite{Lin.2018n04, Earnest.2018lcs} but at the expense of limited coherence caused by flux noise.

Even though the first Fluxonium paper reported data of a qubit operated at zero flux~\cite{Manucharyan.2009qb1}, this bias regime remains largely unexplored since, in recent Fluxonium experiments~\cite{Nguyen.2019}, the first excited state at zero flux was of the plasmon type. However if $\omega_\text{p} = \sqrt{8E_\mathrm{J}E_\mathrm{C}}\gg 2\pi^2 E_\text{L}$ two parabolas corresponding to persistent-current states differing by $\Delta m =2$ anti-cross at zero flux at a lower energy than the plasmon state (see Fig.~\ref{qps_&_fluxonium}). These fluxons are coupled by double QPS via  $E_{S,2}$~\cite{Kalashnikov.2020} and by second order tunneling via $E_{S,1}$. These two mechanisms allow the tunneling of two fluxons or bi-fluxon tunneling and define the the splitting $\Delta_{4\pi}$ between the first and second excited states of the fluxonium, which are respectively the anti-bonding and bonding combinations of two fluxons states differing by $\Delta m =2$. They are particularly interesting from a quantum information perspective since their wave-functions are localized at $\varphi = \pm 2\pi$ in phase representation and are then disjoint from the ground state localized at $\varphi = 0$ (see Fig.~\ref{qps_&_fluxonium}.c). This bi-fluxon tunneling provides then a mechanism to protect the fluxonium qubit against depolarization. In addition to this $T_1$-protection, operating the fluxonium at zero flux offers first order protection against dephasing owing to a sweet-spot similar to the one found at half-flux. To summarize, the qubit presented in this work benefits from an exponential protection in energy relaxation and dephasing coming from charge noise and linear protection in dephasing due to flux noise according to the classification set out in~\cite{Gyenis.2021oc}. 

\section{Circuit Implementation}
\begin{figure}
\centering
\includegraphics[width=1\columnwidth]{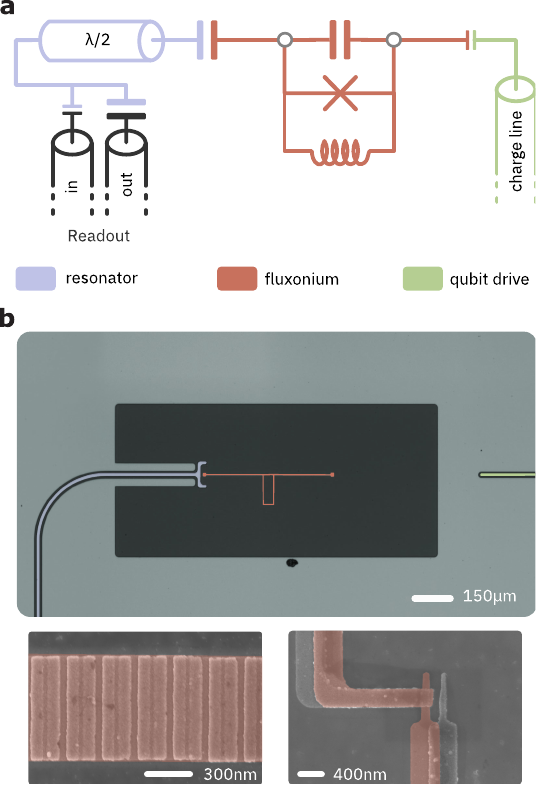}
\caption{\label{implementation} \textit{Device and schematic}. 
\textbf{a.} Electrical circuit of the Fluxonium capacitively coupled to a readout resonator and to a charge line. \textbf{b.} False colored optical image of the sample. Insets show scanning electron microscope (SEM) images of the Josephson junction of the qubit (right) and the Josephson-junction-based inductor (left).}
\end{figure}

\begin{figure}
\centering
\includegraphics[width=1\columnwidth]{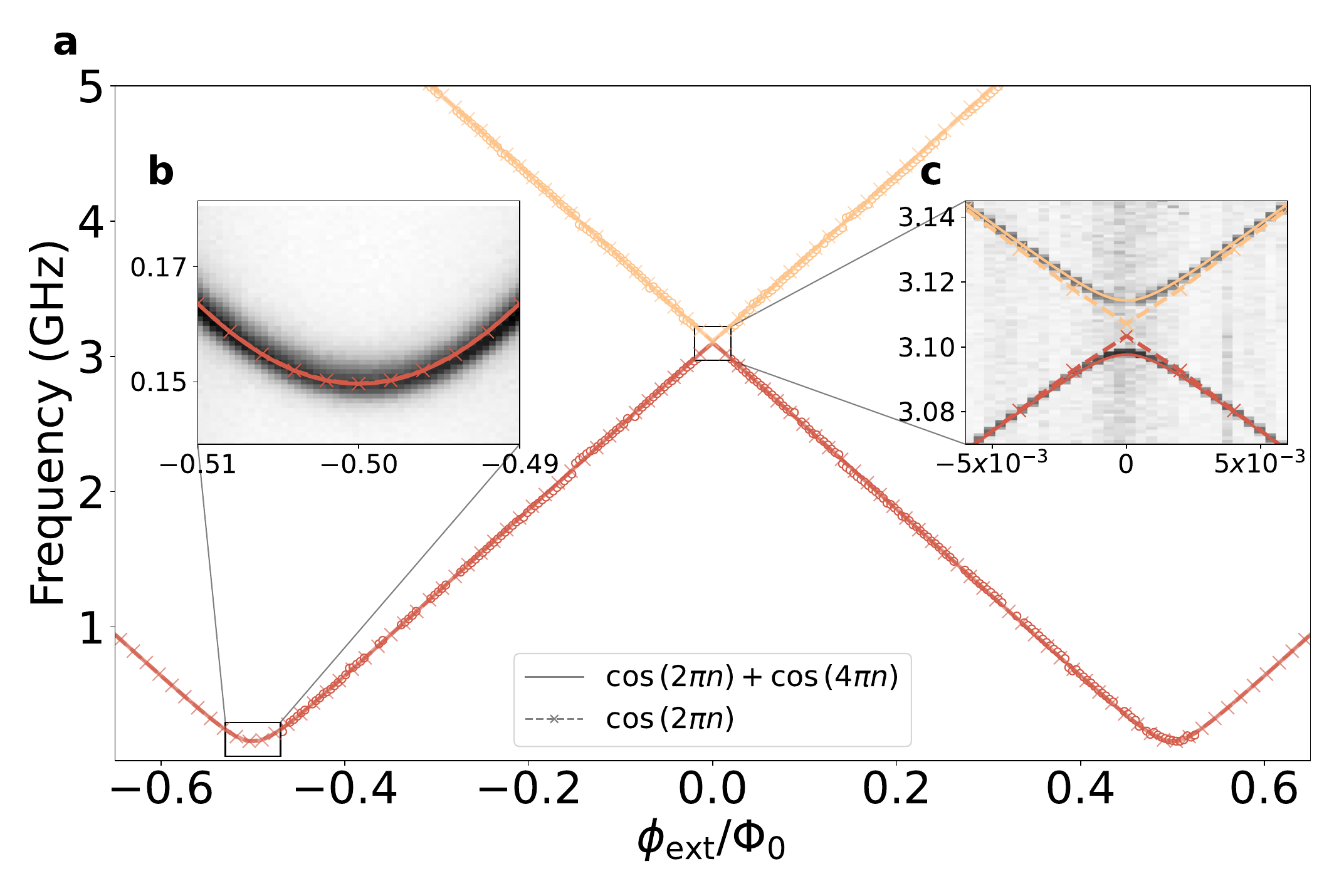}
\caption{\label{spectroscopy} \textit{Sample A spectroscopy}. \textbf{a.} Two-tone spectrum of sample A. The points corresponding to the different transitions are extracted (colored circles) and fitted to both the bifluxon Hamiltonian Eq.~(\ref{dual}) and a simplified version that neglects the $E_{S,2}$ term. The full data as well as the details of the fitting procedures are given in appendix~\ref{spectro_fit}. Measured spectroscopy along with the different fitted models around half integer (\textbf{b.}) and integer (\textbf{c.}) flux quanta.}
\end{figure}

 The fluxonium circuit presented in this work (see Fig.~\ref{implementation}) is a superconducting loop formed by a large super inductance ($\sim 1~$\textmu H) interrupted by a Josephson junction ($E_\text{J}/E_\text{C} \sim 4$). The large inductance is made using a Josephson junction chain of 524 large junctions. The loop is attached to a dipolar antenna, that forms the larger part of the fluxonium capacitance. The antenna is then coupled from one side to a $\lambda/2$ coplanar waveguide resonator with a frequency of $6.89~\text{GHz}$ and a linewidth $\kappa/2\pi \approx 6.5~\text{MHz}$. This represents a rather low external quality factor $Q = 1000$. The qubit is driven by microwave pulses through a charge line that is coupled from the other side of the antenna. The measurement setup is presented in appendix~\ref{setup}. Standard filtering and shielding are used to minimize spurious noise and stray radiation respectively. A Traveling Wave Parametric Amplifier guarantees a good signal-to-noise ratio.

 The two devices presented in this work, labelled A and B, are fabricated with aluminium deposited on a high-resistivity silicon substrate in a planar geometry (see Fig.~\ref{implementation}.\textbf{b}). The fluxonium are fabricated in a single evaporation step, the small junction and the chain are made using the Dolan bridge technique. The first angle evaporation is made at zero degree so that the size of the junctions depends only weakly on resist mask variations. The fabrication process is detailed in the appendix \ref{fab}.

The spectrum of the fluxonium is obtained using conventional two-tone spectroscopy (see appendix~\ref{spectro_fit} and Fig.~\ref{Spec_sampleA}) and is fitted with the fluxonium model given by Eq.~\eqref{H_fluxo} taking into account the coupling to the readout resonator. The obtained parameters are $E_\text{C} = 1.59~\text{GHz}$, $E_\text{J} = 6.02~\text{GHz}$, and  $E_\text{L} = 164.7~\text{MHz}$. The same spectrum was fitted with the dual model given by Eq.~\eqref{dual} and a simplified dual model that neglects the double quantum phase slips amplitude $E_\text{S,2}$ (see Fig.~\ref{spectroscopy}). The fitting procedure results in $E_\text{S,1} = 153~\text{MHz}$, $E_\text{S,2} = 12.98~\text{MHz}$ and $\text{E}_\text{L}^\star = 157 ~\text{MHz}$. Where $\text{E}_\text{L}^\star$ is the renormalized inductive energy due to interband coupling (see appendix~\ref{Bloch}). This clearly shows that the bifluxon tunneling is needed to explain the measured $\Delta_{4\pi}$ and is actively contributing at the zero-flux sweet spot.

\section{Coherence time.}

\begin{figure}
\centering
\includegraphics[width=1\columnwidth]{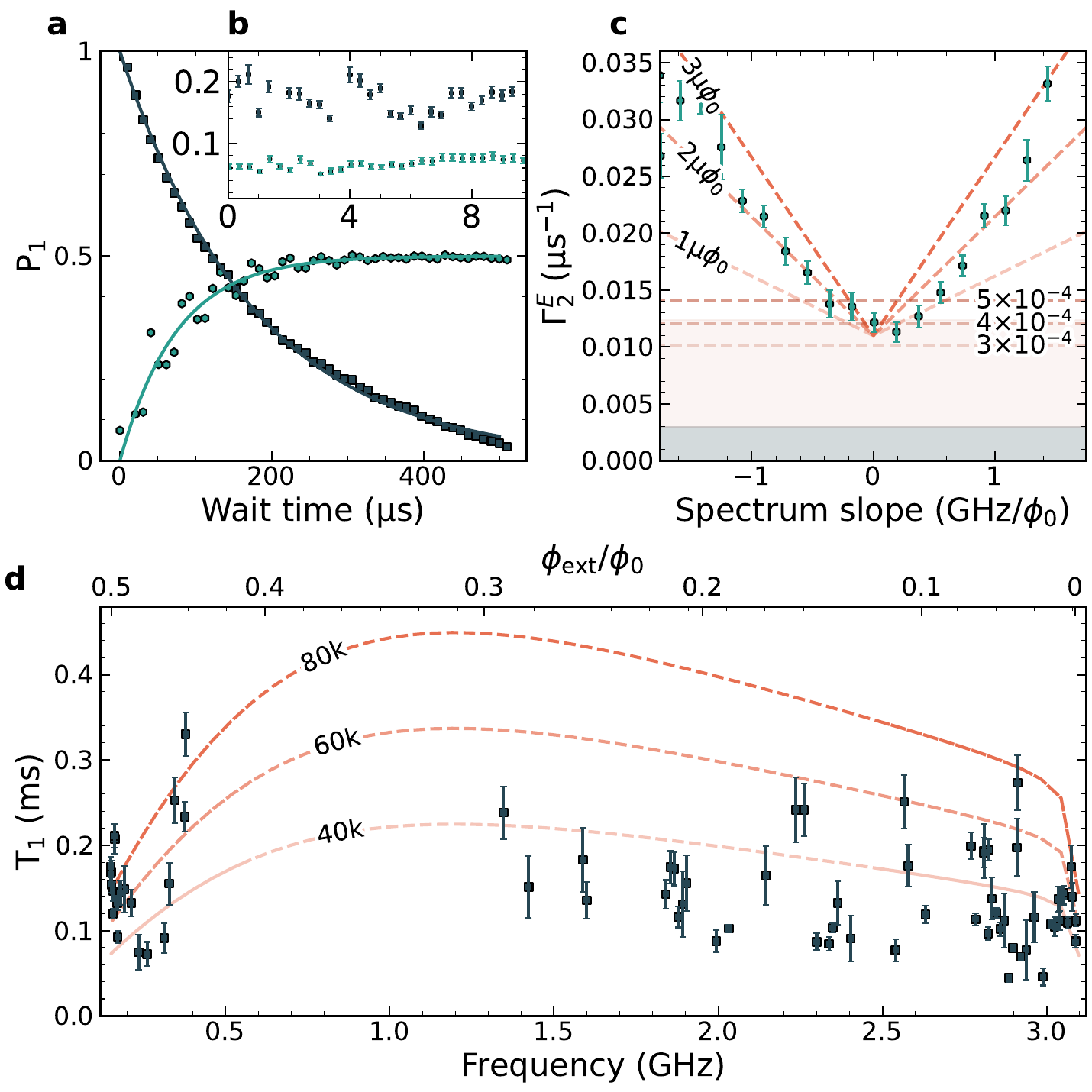}
\caption{\label{Coherence} \textit{Coherence of sample A} \textbf{a.} Interleaved energy relaxation time $\text{T}_1$ (dark teal square marks) and echo coherence time $\text{T}_{2}^E$ (light teal hexagonal markers) measurements averaged over 11 hours. \textbf{b.} Individual measurements corresponding to the panel \textbf{a}. The y-axis represents the corresponding coherence times in milliseconds while the x-axis is time in hours. The energy relaxation $T_1$ remains between $150$ and $220~\mathrm{\mu s}$ while the coherence time $T_2^E$ is between $60$ and $90~\mathrm{\mu s}$ for the entire duration of the measurement.
\textbf{c.} Measured echo coherence time $\text{T}_{2}^E$ as function of flux (light teal hexagonal markers), red dashed lines are obtained for different values of 1/f noise amplitude Eq.~(\ref{1overfdephasing}).  The grey area at the bottom is $\Gamma_1/2$. The orange areas between this latter and the horizontal dashed lines correspond to dephasing from various residual photon numbers. Error bar are one standard deviation.
\textbf{d.} Measured energy relaxation time as function of external flux (dark teal square marks). Near half flux the measurement was performed using a large microwave drive in the beginning of the experiment to reduce the qubit thermal population, while all the other points were measured without this pulse. Dashed lines represent the dielectric loss theory for different dielectric quality factors Eq.~(\ref{capacitiveloss}). }
\end{figure}

Using standard relaxation measurement, we measured a depolarisation time $T_1 = 177 \pm 4~\text{µs}$ at zero flux for sample A. This measurement was obtained by repeating this relaxation measurement over 11 hours to account for the known slow variation over time of this quantity~\cite{burnett_decoherence_2019}. 
The result is displayed in figure~\ref{Coherence}. To get more insights on the energy relaxation mechanisms, we repeated this measurement as a function of the external magnetic flux. A major limitation on the qubit $T_1$ is capacitive loss whose value can be estimated using: 
\begin{equation}
\label{capacitiveloss}
\frac{1}{T_{1, \mathrm{diel}}} = \frac{\hbar\omega_{01}^2}{4 E_\mathrm{C}Q_\mathrm{cap}(\omega_{01})}\coth\left(\frac{\hbar\omega_{01}}{2k_\mathrm{B}T_\mathrm{eff}}\right) \left|\bra{0}\varphi\ket{1}\right|^2
\end{equation}
where $\omega_{01}$ is the angular frequency of the $\ket{0}-\ket{1}$ transition, $Q_\mathrm{cap}(\omega)=Q_\mathrm{cap}(\omega_\mathrm{r})(\omega_\mathrm{r}/\omega)^\epsilon$ is the quality factor associated with capacitive loss and $T_\mathrm{eff}$ is the temperature of the bath~\cite{sun_characterization_2023}. Using this formula with $\epsilon=0.2$, $\omega_\mathrm{r}=2\pi\times 6$~GHz and $T_\mathrm{eff}=20$~mK, which are typical values for this loss channel~\cite{sun_characterization_2023, Nguyen.2019}, we estimate $Q_\mathrm{cap}(\omega_\mathrm{r})\sim6\times10^4$.

The coherence time of sample A is measured using a single $\pi$-pulse echo sequence in order to eliminate low frequency drifts during the measurement.
 The latter was also measured over the course of 11 hours, interleaved with $T_1$ measurements, and was estimated to be $T_2^\text{E} =75 \pm 6\text{ µs} $ at zero flux (see Fig.~\ref{Coherence}\textbf{a}). A similar investigation was conducted around half-flux (see appendix~\ref{half-flux}).
Then, $T_2^\text{E}$ was measured around zero flux to investigate the impact of flux noise on the qubit coherence. 
At integer flux values, we observe that the qubit's coherence exhibits an exponential decay function. Away from the sweet spots, the decay function becomes Gaussian, indicative of $1/f$ flux noise. To estimate the noise amplitude, we utilize the Gaussian pure dephasing rate, given by \cite{Braumuller_2021}:
\begin{equation}
\label{1overfdephasing}
\Gamma_\phi^\text{E} = \sqrt{A_\phi \ln{2} } \left| \frac{\partial \omega}{ \partial \Phi_\text{ext} }\right|
\end{equation}
It is assumed here that the accumulation of the qubit phase follows Gaussian statistics, with a noise power spectral density $\text{S}_\phi (\omega) = A_\phi / \left| \omega \right|$, where $\sqrt{A_\phi}$ represents the flux noise amplitude. The result is reported in Fig.~\ref{Coherence}\textbf{c} where we see that the flux noise seems to be limiting $T_2^\text{E}$ away from sweet spot. From this measurement we estimate $\sqrt{A_\phi} \sim 2$~\textmu$\phi_0$ which is comparable to the state of the art. However, at zero flux another mechanism saturates $T_2^\text{E}$. Flux noise contributes to second order according to $\Gamma_{\phi}^{\text{E}} =\frac{\pi}{2} A_\phi \left| \frac{\partial^2 \omega}{ \partial^2 \Phi_\text{ext}}\right|$~\cite{ithier_decoherence_2005}. However, the noise amplitude reported above would translate into $T_\phi^E \sim  35 ~ \text{ms}$. Another limitation could be the presence of the second excited level. This nearby level induces a dephasing given by $\Gamma_\phi=\Gamma_{12}^{\uparrow}/2$ (see appendix \ref{f-dephasing}). Using the same $Q_\mathrm{cap}$ as above, we obtain $T_\phi=1.5$ ms. Then the most likely hypothesis would be a number of residual thermal photons $\bar n_\mathrm{th}$ in the readout resonator. In the $\bar n_\mathrm{th}\ll 1$ limit we have~\cite{wang_cavity_2019}: 
\begin{equation}
\label{thermalphotondephasing}
    \Gamma_\phi^\text{th}=\frac{ \bar{n}_\text{th}\kappa \chi_{01}^2}{\kappa^2+\chi_{01}^2}
\end{equation}


Our measure of $T_2^\text{E}\sim 75~\mu \text{s}$ corresponds to $\bar{n}_\text{th}= 4\times 10^{-4}$, which converts to an effective temperature $T_\text{eff} \approx 42$ mK. This temperature is in agreement with the values reported in the literature~\cite{Somoroff.2023, yan_distinguishing_2018}. Similar measurements were conducted for sample B. They are presented in appendix~\ref{SampleB} and summarized in table~\ref{tab:sample_Table}.

\begin{table*}
    \centering
    \begin{tabular}{|c || c|c|c|c|c|c|c|} \hline 
          &  $\text{E}_\text{J}$ (GHz) & $\text{E}_\text{C}$ (GHz) & $\text{E}_\text{L}$ (MHz)& $\text{E}_\text{s,1}$ (MHz) & $\text{E}_\text{s,2}$ (MHz) & $\text{E}_\text{L}^\star$ (MHz) & \# JJ in chain  \\  \hline
         Sample A & $6.01$ & $1.59$ & $165$ & $153$ & $13$ & $157$ & $524$ \\ \hline
         Sample B & $5.76$ & $1.62$ & $162$ & $176$ &  $17$ &  $154$ & $524$ \\ \hline 
         & $\text{T}_\text{1} (\phi_\text{ext} = 0) $ ($\mu$s) &$Q_\mathrm{cap}$ & $\langle 0\lvert \hat{n}\rvert 1\rangle (\phi_\text{ext} = 0) $ &$\text{T}_\text{2}^\text{E} (\phi_\text{ext} = 0) $ ($\mu$s) & $\text{f}_\text{r} $ (GHz)  & $\kappa_\text{r} $ (MHz) & $\chi_{01}$ (MHz)   \\ \hline
         Sample A& $177.3$ & $6\times10^{4}$ & $0.063$ & $74.6$ & $6.908$ & $6.5$ & $6.8$\\ \hline
         Sample B& $87.6$ & $4\times10^{4}$ & $0.071$ & $45.7$ & $6.89$ & $5.4$ & $6.9$\\ \hline 
    \end{tabular}
    \caption{Sample parameters and coherence times.}
    \label{tab:sample_Table}
\end{table*}

\section{Discussion}


First, compared to the standard implementation of the fluxonium, our qubit frequency is in a frequency regime where $f\gg k_\mathrm{B}T/h$. Hence, this qubit does not need to be initialized prior to the measurements.

Moreover, we would like to stress that our circuit is very accurately described by the effective QPS Hamiltonian Eq.~(\ref{dual}). Also it appears clearly that the anti-crossing between the 01 and 02 transitions at zero flux cannot be described quantitatively without taking into account the $\cos (4\pi \hat{n})$ term. This constitutes the first important result of our experiment: using a relatively simple circuit design, we could obtain a large bi-fluxon tunneling amplitude ($E_\text{S,2} \sim 13~\text{MHz}$). This value is smaller than the one obtained in the initial work of Kalashnikov et al.~\cite{Kalashnikov.2020} and we do not have the ability to tune in-situ the ratio $E_\text{S,2}/E_\text{S,1}$, but our qubit is a single mode circuit and does not require any electrostatic gate tuning. It is thus  insensitive to slow charge drifts, quasi-particle poisoning and much easier to fabricate and model, which are very important features when high coherence is required.

Our circuit also enables us to explore the dynamics of a Josephson junction placed in a high-impedance environment, and more specifically the physics associated with Bloch bands~\cite{Likharev.1985}. It is important to stress that the very small $E_\text{L}$ regime was reached by the Blochnium~\cite{Pechenezhskiy.2020} but in the experiment by Pechenezhskiy et al. the low energy states are localized inside the first Bloch band ($s=0$) and are plasmons states~\cite{Koch.2009}. As such, they have very low flux dispersion and therefore benefit from very good protection against flux noise, but as their wave functions are close to those of the harmonic oscillator, they have no protection against bit flips. In contrast, the quantum states of our qubit are persistent-current states coupled by multiple QPS. This allows us to use bifluxon tunneling and encode quantum information in fluxon number parity. Another interesting point about Bloch band physics is that the inductive energy $E_\text{L}^*$ of the dual model Eq.~(\ref{dual}) is lower than that found by the exact model Eq.~(\ref{H_fluxo}). This slight reduction of the inductive energy can be understood as an interaction between the low-energy persistent-current states and the plasmon state bound to the second band ($s=1$), as discussed in appendix~\ref{Bloch}.

We are now turning our attention to the potential evolution of this qubit and to ways of improving it. It is then useful to understand the parametric of the spectroscopic gaps $\Delta_{2\pi}$ and $\Delta_{4\pi}$. They are given by~\cite{Vakhtel.2023}:
\begin{equation}
\label{equation: E2PI}
\Delta_{2\pi}= 4\omega_\text{p} \left(\frac{2 E_\text{J} }{\pi^2 E_\text{C}} \right)^{\frac{1}{4}}  e^{-\sqrt{\frac{8 E_\text{J}}{E_\text{C}}}  + \frac{14 \zeta(3) E_\text{L}}{\omega_\text{p}}},
\end{equation}
with $\zeta$ the Riemann zeta function and~\cite{Vakhtel.2023}:
\begin{equation}
\label{equation: E4PI}
\Delta_{4\pi}=\frac{\Delta_{2\pi}^2}{4\pi^2 E_\text{L}}\left(8 \sqrt{\frac{8 E_\text{J}}{E_\text{C}}}\right)^{\frac{2E_\text{L}\pi^2}{\omega_{\text{p}}}}.
\end{equation}
The equations  \eqref{equation: E2PI} and \eqref{equation: E4PI} are valid in the semiclassical WKB approach where $E_\text{C} \ll E_\text{J} $ and  $\text{max}(\Delta_{2\pi}, \Delta_{4\pi}) \ll E_{\text{L}} \ll \omega_{\text{p}}= \sqrt{8 E_\text{J} E_\text{C}}$. They actually provide very good approximate values to our experimental findings as shown in table~\ref{Deltas} and can thus be used to optimize further the properties of this partially protected fluxonium. Knowing that the aim is to obtain a high $\Delta_{4\pi}$ value while keeping $\omega_{\text{p}}$ high, it is clear that we should aim for $E_{\text{L}}$ small and decrease the $E_\text{J}/E_\text{C}$ ratio. However, we must also be careful not to reduce $E_\text{J}/E_\text{C}$ too much, so that fluxonium is not too light, which would cause it to enter the Blochnium regime and remove its parity protection.

\begin{table}
\centering
\begin{tabular}{| m{1cm} | m{2.5cm}| m{2.5cm} | }
 \hline
   & Experiment & WKB formula  \\
 \hline
 $\Delta_{2\pi}$ & $153\text{ MHz}$ & $184\text{ MHz}$ \\
 \hline
 $\Delta_{4\pi}$ &  $13\text{ MHz}$ & $21\text{ MHz}$  \\
 \hline
\end{tabular}
\caption{\label{Deltas} Comparison between the experimental data and WKB formula \eqref{equation: E2PI} and \eqref{equation: E4PI}.}
\end{table}

The $T_1$ flux dependence (see Fig.~\ref{Coherence}\textbf{d}) is compatible with a loss tangent of $1/Q_\mathrm{cap}\sim 10^{-5}$, which is more than an order of magnitude higher than the state of the art. An obvious improvement for this qubit would be an optimized fabrication process involving enhanced cleaning steps or better quality materials. Combining this with the optimization of parameters $E_\text{J}$, $E_\text{C}$, $E_\text{L}$ mentioned above, it is reasonable to anticipate lifetimes of several milliseconds. Regarding the dephasing rate, since our qubit is limited by the resonator shot noise, we can easily improve it by either increase the resonator linewidth or reducing the coupling rate. By doing so, the limiting factor should be the dephasing coming from the second excited state. However, it will also be greatly improved with lower capacitive loss. Hence, a dephasing time of several milisecond for a qubit in the gigahertz range should be within experimental reach. Another important point is that there is no transition in the vicinity of the computation transition that can be activated by a microwave excitation. The closest transition is 02 which exhibits no sensitivity to electrical field thanks to the symmetry of the wave-functions (see the vanishing contrast in the spectroscopy in Fig.~\ref{spectroscopy}.\textbf{c}). The next closest transition is 13 which is separated by several hundreds of megahertz from the qubit transition. This qubit is therefore compatible with the implementation of single-qubit gate at least as fast as the ones implemented on Transmons \cite{lazuar2023calibration}.

\section{Conclusion}

 Our work demonstrates a regime of operation of the fluxonium qubit in which the ground and excited state wave-functions belong to different fluxon parity. The qubit relaxation time seems to be limited by capacitive losses. However, while the estimated losses are large compared to the best values reported in the literature, the measured relaxation time is above 100 \textmu s in a 2D geometry with a qubit frequency around 3 GHz. This is possible owing to the built-in protection against energy relaxation stemming from the reduced overlap of the wave-functions of the qubit. Similarly to the conventional regime of operation for Fluxonium, our qubit is immune to low-frequency charge noise and has a first-order protection against flux noise. We thus anticipate that these decoherence mechanisms should not limit the qubit coherence until it reaches the $\sim 10$ ms range. The anharmonicity is larger than a typical Transmon and since the qubit frequency is in the gigahertz range it does not suffer from thermal effects, as it is often the case with the usual Fluxonium operated at half flux.

 Despite large dielectric losses, which can readily be improved using state-of-the-art fabrication techniques, our first implementation of a Fluxonium qubit protected by bi-fluxon tunneling is already providing good coherences without requiring a very stringent parameter regime. Moreover, our approach is not only interesting from a quantum information point of view, it could open exciting avenue for waveguide-QED and more particularly for experiments investigating Bound-state In a Continuum~\cite{Hita-Pérez.2022}.
    
\begin{acknowledgments}

This work is supported by the European Union’s Horizon 2020 research and innovation program under grant agreement no. 101001310, by Horizon Europe programme HORIZON-CL4-2022-QUANTUM-01-SGA via the project 101113946 OpenSuperQPlus100 and by the French grant ANR-22-PETQ-0003 under the ‘France 2030 plan’. T.V.
has received funding from the European Research Council
(ERC) under the European Union’s Horizon 2020 research
and innovation programme. The sample was fabricated in the clean room facility of Institute Neel, Grenoble. We sincerely thank all the clean room staff for help with fabrication of the devices. We would like to acknowledge E. Eyraud for his extensive help in the installation and maintenance of the cryogenic setup and J. Minet for significant help in programming the high-speed pulse generation and data acquisition setup. We also thank J. Jarreau, D. Dufeu and L. Del Rey for their support with the experimental equipments.
We are grateful to S. Deléglise for lending us a first sample and to D. Basko, Z. Leghtas and E. Flurin for insightful discussions regarding this project. We thank the members of the superconducting circuits group at Neel Institute for helpful discussions.\\
\end{acknowledgments}

\appendix

\section{Bloch Band decomposition}
\label{Bloch}

In this appendix, we discuss in more details the link between dual Hamiltonian Eq.~(\ref{dual}) and the fluxonium Hamiltonian Eq.~(\ref{H_fluxo}). We will show how the formalism developed in~\cite{Koch.2009} can be used to explain how the inter-band coupling both renormalize the inductive energy $E_\mathrm{L}^\star$ and induce the plasmon states observed in the energy spectrum of Fig.~\ref{spectroscopy}. 

First, we diagonalize the $4 E_\mathrm{C}\hat{n}^2-E_\mathrm{J}\cos\hat\varphi$ part of the Hamiltonian. We call the basis where it is diagonal $\ket{n,s}$, where $n$ stands for the quasi-charge while $s$ stands for the band index. In the basis of fluxons, or equivalently flux quanta, $ |m, s\rangle=\int_{-\frac{1}{2}}^{\frac{1}{2}} dn e^{-i 2 \pi m n}|n, s\rangle$, the Hamiltonian is:
\begin{align}
    \hat H_{\mathrm{f}}&=2 \pi^2 E_{\mathrm{L}}\left(\sum_{m,s}m\ket{m,s}\bra{m,s}+\frac{\Phi_{\mathrm{ext}} }{\Phi_0} +\frac{\hat{\Omega}}{2\pi}\right)^2+ \nonumber\\ 
    &+ \sum_{k=0}^\infty\sum_{m,s}\left(\frac{E_{s,k}}{2} \ket{m+k,s}\bra{m,s} +h.c.\right) \label{eq:dual_app}
\end{align}
where $E_{s,0}$ is equal to $ s\omega_p= s\sqrt{8E_J E_C}$ plus some exponentially small contributions that depend on the band index and $\hat\Omega$ is the inter-band coupling operator, expressed as
\begin{align}
 \bra{p,r}   \hat{\Omega} \ket{m,s}   = \int^{1/2}_{-1/2} dn  \ \Omega^{r,s}(n)e^{-i2 (m-p) \pi n} \label{eq:ft_dual}
\end{align}
and
\begin{align}
    \Omega^{r,s} (n) = i \int_{0}^{2\pi} d \varphi u^{*}_{r,n} (\varphi)\partial_{n} u_{s,n}(\varphi)
\end{align}
where $e^{in\phi}u_{r,n}(\varphi) = \braket{\varphi|n,r}$ and  $u_{r,n}(\phi+2\pi)= u_{r,n}(\phi) $ are normalized such that $\int^{2\pi}_{0} u^{*}_{s,n} u_{r,n} d\phi =\delta_{r,s}$. 
In the harmonic $E_C \ll E_J$ limit the wave-functions in the phase space become infinitely localized and we have
\begin{align}
    \Omega^{s,r}(n)=\left(2 E_\mathrm{C} / E_\mathrm{J}\right)^{1 / 4}\left(\sqrt{s} \delta_{s, r+1}+\sqrt{r} \delta_{s+1, r}\right).\label{eq:inter_band_app}
\end{align}
This gives us coupling that is diagonal in flux. Hence, for the lowest band $s=0$ and using Eq.~(\ref{eq:dual_app}), (\ref{eq:ft_dual}) and (\ref{eq:inter_band_app}), we find that the coupling matrix element at lowest order in $E_\mathrm{C}/E_\mathrm{J}$ is given by $V_{0,1}= 2\pi( m+\Phi_{ext}/\Phi_0) E_{L} \Omega^{0,1} $. The virtual transition induces a contribution to the kinetic energy for the lowest band:
\begin{align}
    H_{kin}^{0} &\approx 2\pi^2 E_L \left(m+\frac{\Phi_{ext}}{\Phi_0}\right)^2- \frac{V_{0,1}^2}{\omega_p} \\
    &= 2\pi^2 E_L \left(1-\frac{E_L}{E_J} \right)\left(m+\frac{\Phi_{ext}}{\Phi_0}\right)^2
\end{align}

\begin{figure}
\centering
\includegraphics[width=0.8\columnwidth]{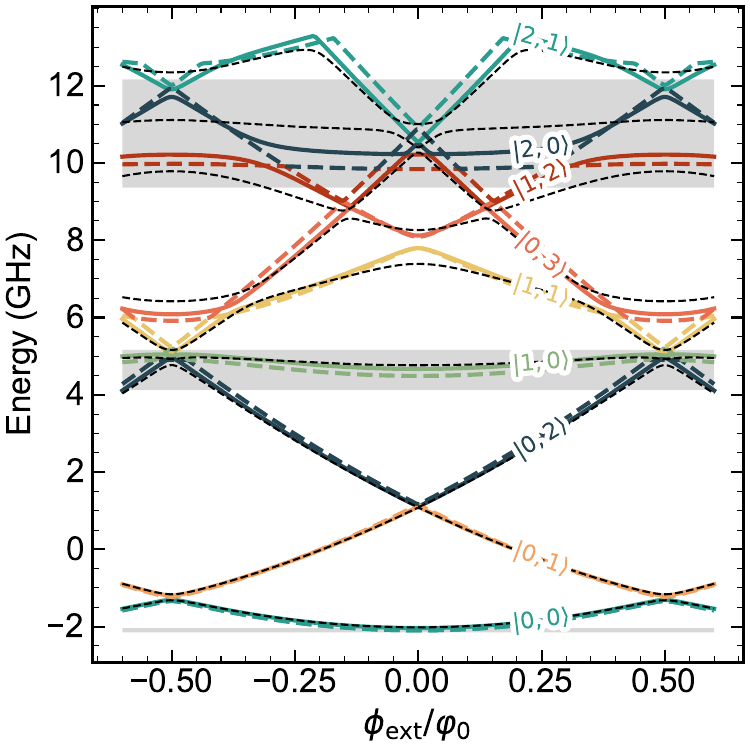}
\caption{\textit{Spectrum of the fluxonium and its dual model.} The black dashed lines are given by the exact model. The dashed line is the dual model without interband coupling while the plain line are accounting for it.}
\label{Spec_sampleA_theory}
\end{figure}

Hence $E_\mathrm{L}^\star = E_\mathrm{L}(1-\frac{E_\mathrm{L}}{ E_\mathrm{J}})$, explaining how the inductive energy can be renormalized by the inter-band coupling. This renormalization effect can be seen in Fig.~\ref{Spec_sampleA_theory}, where we can see that the parabolas formed by the lowest energy levels are slightly more flared in the model that takes inter-band coupling into account than in the one that does not. 

In addition, a second striking effect of taking into account the upper bands of the dual model is that it enables the reproduction of energy levels with low flux dependence, corresponding to the plasmon-type transition. In fact, these transitions can be understood as inter-band transitions in the dual model.  
We can see that these energy levels are not present in Fig.~\ref{qps_&_fluxonium}\textbf{e}, where the upper bands are neglected, whereas they are produced when they are taken into account.

\section{Sample fabrication}
\label{fab}

A high-resistivity Si(100) wafer (10 k$\Omega.$cm), was treated for 30 seconds with a 4\% HF solution to remove its native oxide layer. Approximately 150 nm of Aluminum (Al) was deposited at 0.5 nm/sec by e-gun evaporation system (Plassys). The resonator and feedline were patterned using 80 keV e-beam lithography (NanoBeam Ltd.), followed by wet etching (Aluminium etchant-Type A from Transene Inc.). 

The Josephson junctions were fabricated using the standard Dolan bridge technique. First, the PMMA/MMA (ARP 617.14/600.07) was spin coated at 1500 rpm (with a 1500 rpm/$m^2$ acceleration) for 30 seconds. The wafer was subsequently baked at 200 $\tccentigrade$ for 10 minutes. Following this, a second layer of PMMA 4\% (ARP.679.04) was coated at 5000 rpm (with a 5000 rpm/m$^2$ acceleration) for 30 seconds and then baked at 180 $\tccentigrade$ for 5 minutes. The junctions were patterned using the same 80 keV e-beam system with a current of 1 nA. The wafer was developed in a cold ($\approx 1.3 \tccentigrade$) solution of $H_2O$/IPA (1:3) for 60 seconds, followed by a rinse in water ($H_2O$) for 60 seconds. After the cold development step, a gentle $O_2$ RIE (Reactive Ion Etching) was carried out at a power of 10 W and a pressure of $7.0\times10^{-2}$ mbar.


Aluminium deposition:
\begin{enumerate}[I]
\item The evaporation chamber was pumped for 15 hours before deposition, reaching a final pressure of $2.0\times10^{-8}$ mbar. To further reduce chamber pressure, a Ti evaporation step was employed, resulting in a pressure of $1.2\times10^{-8}$ mbar. A total of 30 nm of Ti was deposited at a rate of 0.1 nm/sec, with the shutter avoiding the deposition on the wafer.

\item A the first evaporation of 20 nm of Al at a rate of 0.1 nm/sec was done with an angle of 0 degrees. The pressure in the chamber during deposition was $2.6\times10^{-8}$ mbar.

\item Subsequently, a static oxidation process for Al was carried out. 200 sccm of $O_2$ were introduced in the chamber until the pressure reached 5 mbar for about 5 minutes. The loadlock was pumped down to $1.1\times10^{-6}$ mbar before opening the valve to the main chamber.

\item Another Ti evaporation was performed to further decrease the chamber pressure to $2.8\times10^{-8}$ mbar, depositing a total of 30 nm of Ti at a rate of 0.1 nm/sec with the shutter, avoiding the deposition on the wafer.

\item The sample holder was rotated to 17.5 degrees and 50 nm of Al was deposited at a rate of 0.1 nm/sec, with the pressure during deposition at $3.0\times10^{-8}$ mbar.
\end{enumerate}

The lift off was done in a NMP solution at a temperature of 80$\tccentigrade$. Subsequently, the wafer was rinsed successively for 30 seconds in three separate beakers containing acetone, ethanol, and IPA. Finally, the wafer was dried with $N_2$.

\section{Experimental setup}
\label{setup}

\begin{figure}
\centering
\includegraphics[width=1\columnwidth]{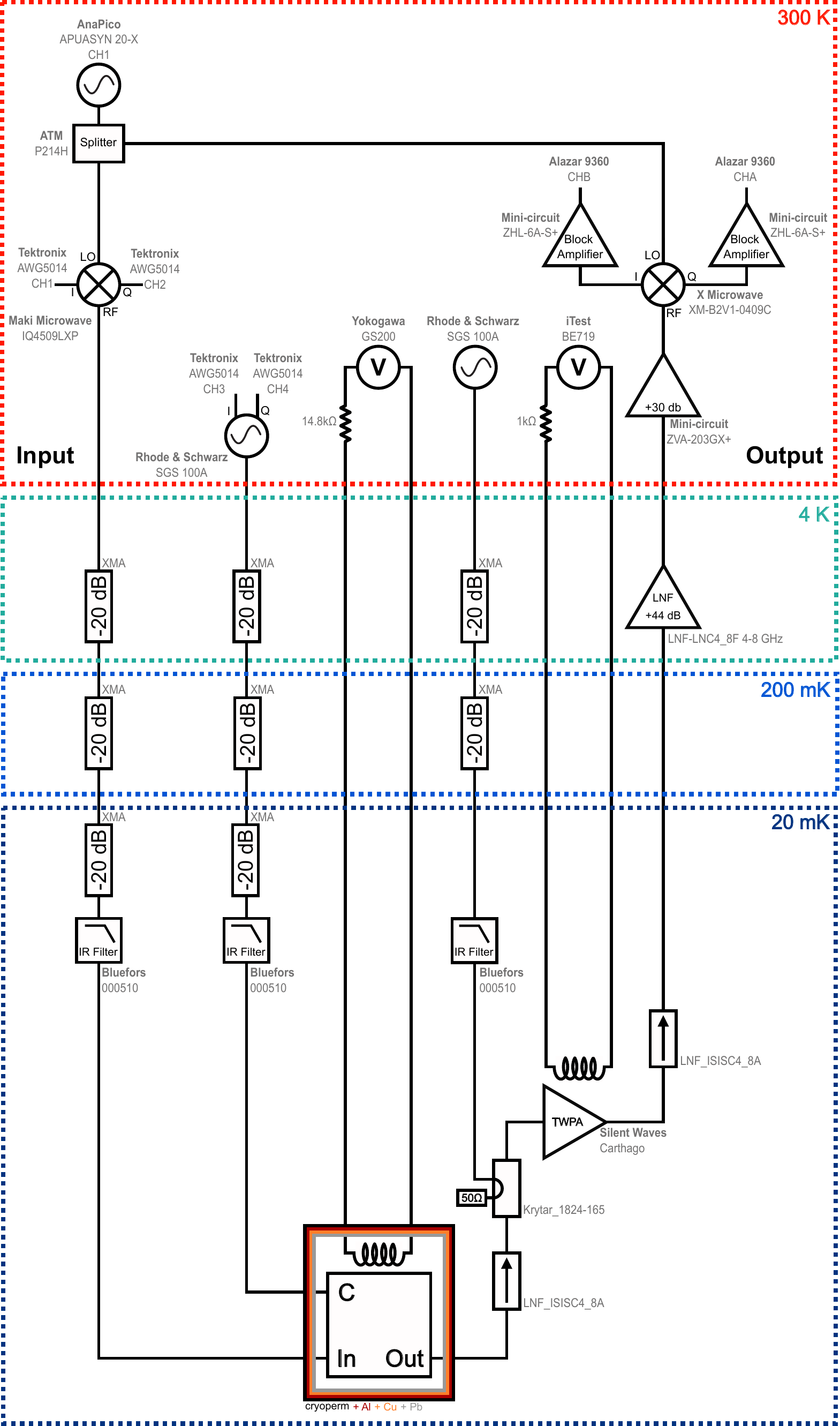}
\caption{\textit{Room temperature and cryogenic wiring schematic of the experimental setup}. }
\label{wiring}
\end{figure}

The experiments were conducted in a Sionludi dilution refrigerator, maintaining a base temperature of 20 mK. A detailed wiring schematic of the experiment is illustrated in Fig.~\ref{wiring}. To mitigate thermal noise, we implemented 40 dB attenuation on the TWPA (Silent Waves Carthago) pump line and 60 dB attenuation on both the readout input line and the charge line. Additionally, high-frequency noise is filtered out at the mixing chamber using Bluefors 000510 IR filters. The sample holder is encased within a copper shield attached to the cryostat's mixing chamber, further protected by a layer of mu-metal and two layers of superconducting shields (Al and Pb).

Pulse generation is accomplished using a Tektronix AWG. The readout signal produced by the AWG is up-converted at room temperature via a Marki IQ mixer, employing AnaPico APUASYN 20-X as the Local Oscillator (LO) source for the input line. For the qubit drive line/charge line (marked as \textbf{C} in Fig.~\ref{wiring}), we utilize the internal IQ mixing stage of the Rhode and Schwarz SGS 100A, with the I and Q ports connected to the Tektronix AWG. The TWPA pump is similarly generated using the Rhode and Schwarz SGS100A. This pump signal is directed to the TWPA through the -20 dB port of a Krytar$\_$1824-165 directional coupler (which makes overall -60 db of attenuation for this signal before reaching the TWPA), located just before the TWPA at the mixing chamber. 

The output signal undergoes initial amplification by the TWPA at the mixing chamber, followed by further amplification using a LNF 4-8 amplifier at the 4K stage. To prevent leakage of the TWPA pump signal and to reduce amplifier noise at the 4K stage from impacting the sample, dual junction LNF isolators are employed: one between the TWPA and the sample, and another between the TWPA and the LNF amplifier at the 4K stage. At room temperature, the signal receives additional amplification through a Mini-circuit ZVA-203GX+ before down-conversion using an X Microwave IQ mixer, with AnaPico APUASYN 20-X serving as the LO source. The two quadratures of the signal, I and Q, are further amplified by a Mini-circuit ZHL-6A-S+ DC gain block, before being fed into an Alazar acquisition card. The global flux coil for both the qubit and the TWPA is DC biased using the Yokogawa GS200 and iTest BE719, respectively.

\section{Spectroscopy fitting procedure}
\label{spectro_fit}
To match the spectrum of the circuit, we begin by determining an initial guess capable of replicating the observed lines in the spectrum. With this guess, we can identify the lines in the spectrum as transitions between states $\lvert i \rangle$ and $\lvert j \rangle$. Subsequently, for a given flux point $\phi_\text{ext}^\text{k}/ \phi_0$, we extract a set of frequencies for transitions $ij$ denoted as $S^\text{k} = \{ f_{ij}^\text{k}\}$. We define a cost function for the set $S^\text{k}$ as:
\begin{align}
    C_S &= \sum \limits_\text{k, i, j} \left\lVert \lambda_j(\text{E}_\text{J}, \text{E}_\text{C},  \text{E}_\text{L}, f_r, g, \phi_\text{ext}^\text{k}/ \phi_0) \right. \nonumber \\
    &\qquad - \left. \lambda_i(\text{E}_\text{J}, \text{E}_\text{C},  \text{E}_\text{L},  f_r, g,\phi_\text{ext}^\text{k}/ \phi_0 ) - f_{ij}^\text{k} \right\rVert
\end{align}

Here, $\lambda_i$ represents the $i^\text{th}$ eigenvalue of the system Hamiltonian (with units normalized to $h = 1$):
\begin{align}
    \hat{H} & = 4 E_\text{C} \hat n^2 - E_\text{J} \cos \hat\varphi+\frac{E_\text{L}}{2}\left(\hat\varphi+\frac{\phi_\text{ext}}{\varphi_0}\right)^2 \nonumber \\
    & + f_r (\hat{a}^\dag \hat{a} + 1/2) + g \hat{n}(\hat{a}^\dag + \hat{a})
\end{align}

Here, $f_r$ is the resonator frequency, and $g$ is the coupling factor between the resonator and the fluxonium. For fitting purposes, the cost function is minimized for the parameters $(\text{E}_\text{J}, \text{E}_\text{C},  \text{E}_\text{L}, f_r, g)$ for a given set $S^\text{k}$. 
To initialize the fitting procedure, we supply the algorithm with an initial parameter guess and a chosen set $S^\text{k}$. The initial parameters are approximated by overlaying theoretical predictions onto experimental data, seeking a rough alignment with most observed lines. Selecting an appropriate set $S^\text{k}$ is pivotal for effective fitting. As we aim to fit five parameters, we require at least five data points. Understanding each parameter's impact on the spectrum helps in this selection: for instance, $\text{E}_\text{L}$ primarily influences the slope of fluxon lines, while the ratio $\text{E}_\text{J}/ \text{E}_\text{C}$ affects the sharpness at the intersection of two coupled fluxon lines. The resonator frequency is usually well-established, aiding in setting this initial guess, and for $g$, we select a point that highlights the interaction between the fluxonium and the resonator.

To validate the fit, we verify if the optimized parameters can accurately replicate all experimental lines. Occasionally, adding specific points to $S^\text{k}$ and repeating the fitting process is necessary to achieve a higher accuracy.

\begin{figure}
\centering
\includegraphics[width=1\columnwidth]{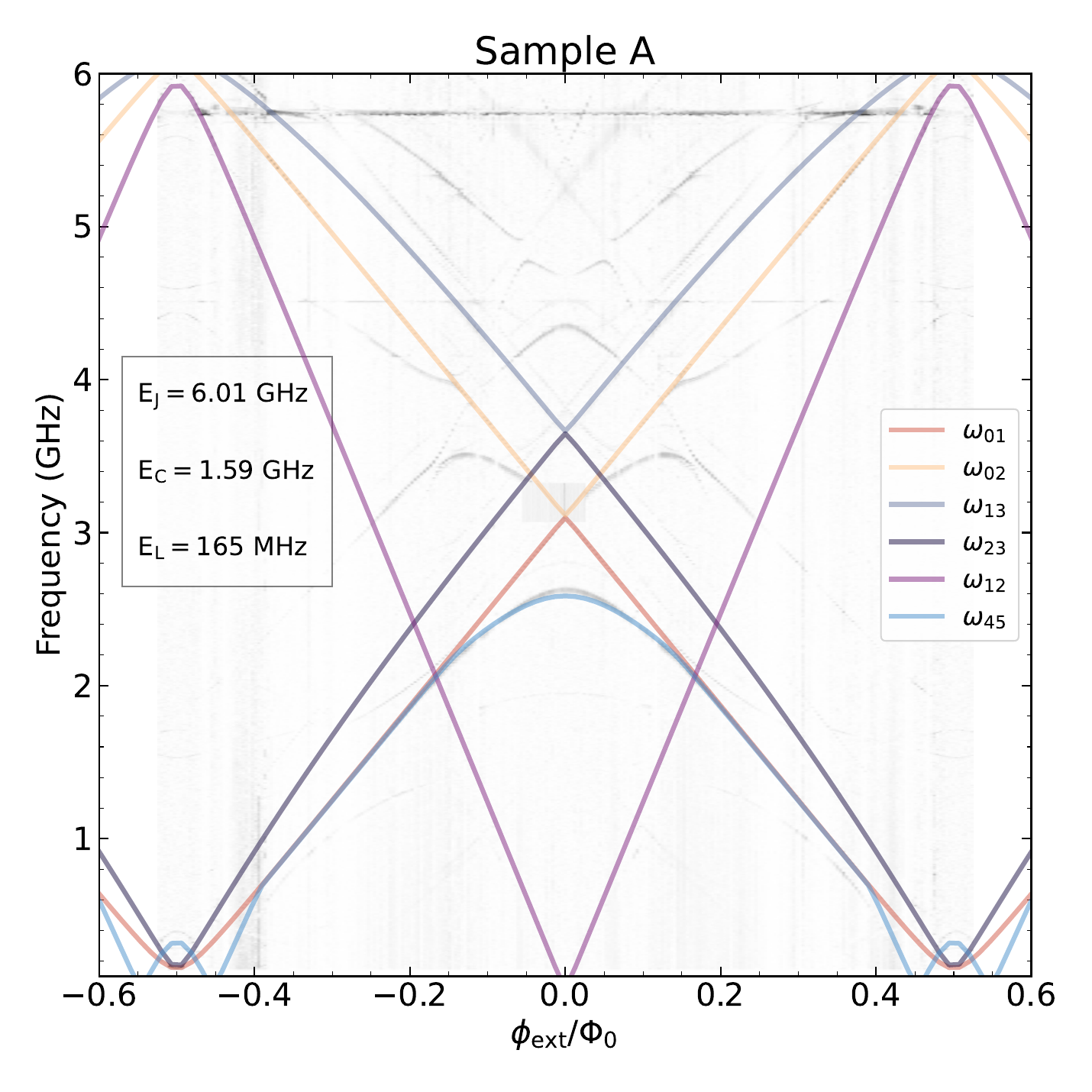}
\caption{\textit{Spectrum of sample A}. Experimental spectroscopy of sample A. The colored lines corresponds to the fit to the Hamiltonian of the fluxonium coupled to the readout resonator.}
\label{Spec_sampleA}
\end{figure}

\section{Qubit coherence at half-integer flux}
\label{half-flux}
\begin{figure}
\centering
\includegraphics[width=1\columnwidth]{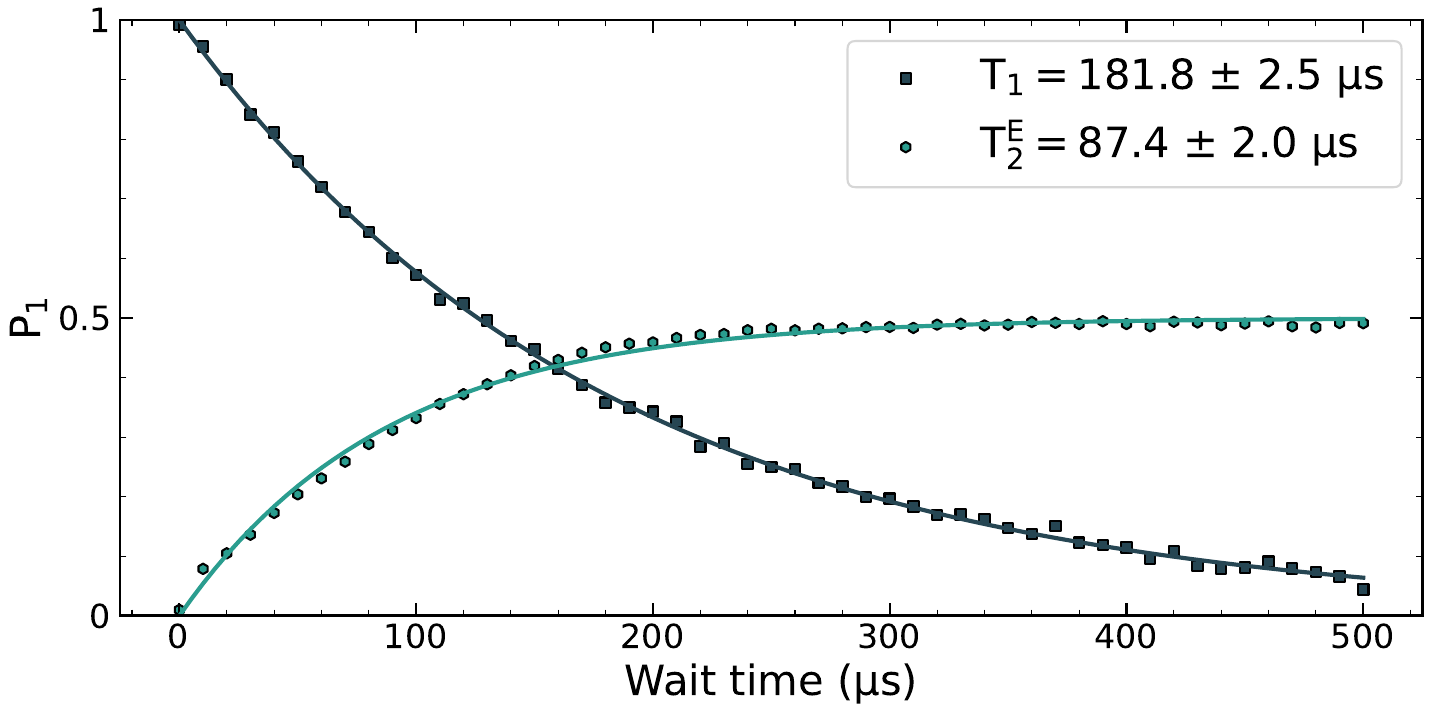}
\caption{\label{Coherence_half} \textit{Coherence of sample A at half-integer flux}. Interleaved energy relaxation time $\text{T}_1$ and echo coherence time $\text{T}_{2}^E$ measurements averaged over 12 hours. }
\end{figure}

In addition to the coherence data presented in the main text, We also characterized Sample A at half-integer flux. The qubit transition frequency at half flux is approximately 150 MHz. To perform a Rabi oscillations, we employ a high-power microwave pulse at the cavity frequency \cite{Ficheux_2021_ZGate} to prepare the qubit. Subsequently, we measure the relaxation time using a standard $\pi$-pulse sequence, obtaining $T_1 = 182~$\textmu s, and the Echo coherence time with a typical echo sequence $T_2^\text{E} = 88$~\textmu s  (see Fig.~\ref{Coherence_half}). The measurements are interleaved and averaged over a 12-hour interval. The qubit on half flux shows approximately the same $T_1$ as on integer flux, but at a significantly lower frequency.
At integer flux, the effective quality factor is $Q_\text{eff} = \omega_{01} \times T_1 = 2\pi \times 3 \, \text{GHz} \times 171 \, \mu\text{s} = 3.2 \times 10^6$, while at half-integer flux, the effective quality factor is more than one order of magnitude lower: $Q_\text{eff} = 2\pi \times 150 \, \text{MHz} \times 182 \, \mu\text{s} = 170 \times 10^3$ due to the large wave-function overlap.


\section{Spectroscopy and coherence times of sample B}
\label{SampleB}
\begin{figure}
\centering
\includegraphics[width=1\columnwidth]{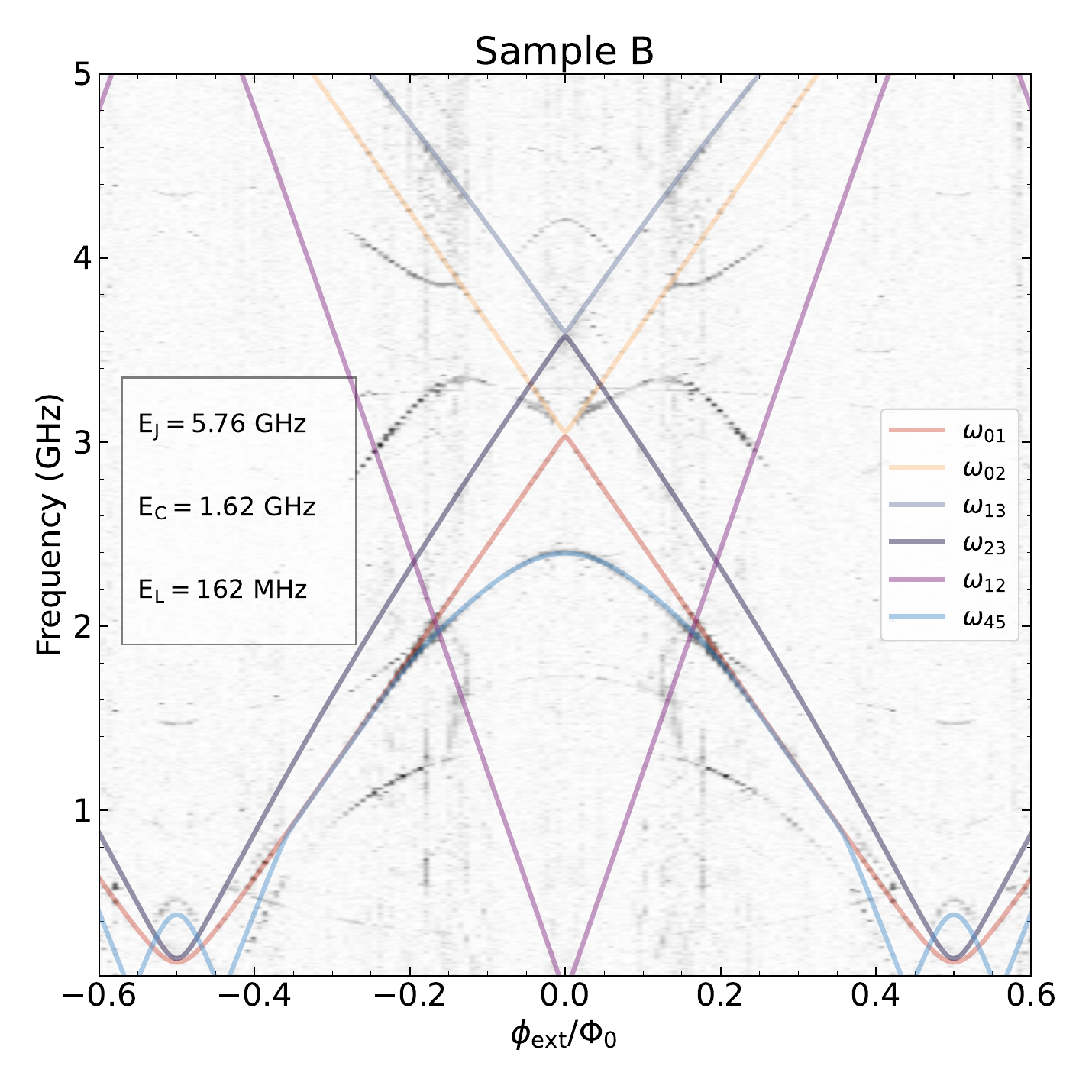}
\caption{\textit{Spectrum of sample B}.  Experimental spectroscopy of sample B. The colored lines corresponds to the fit to the Hamiltonian of the fluxonium coupled to the readout resonator.}
\label{Spec_sampleB}
\end{figure}
\begin{figure}
\centering
\includegraphics[width=1\columnwidth]{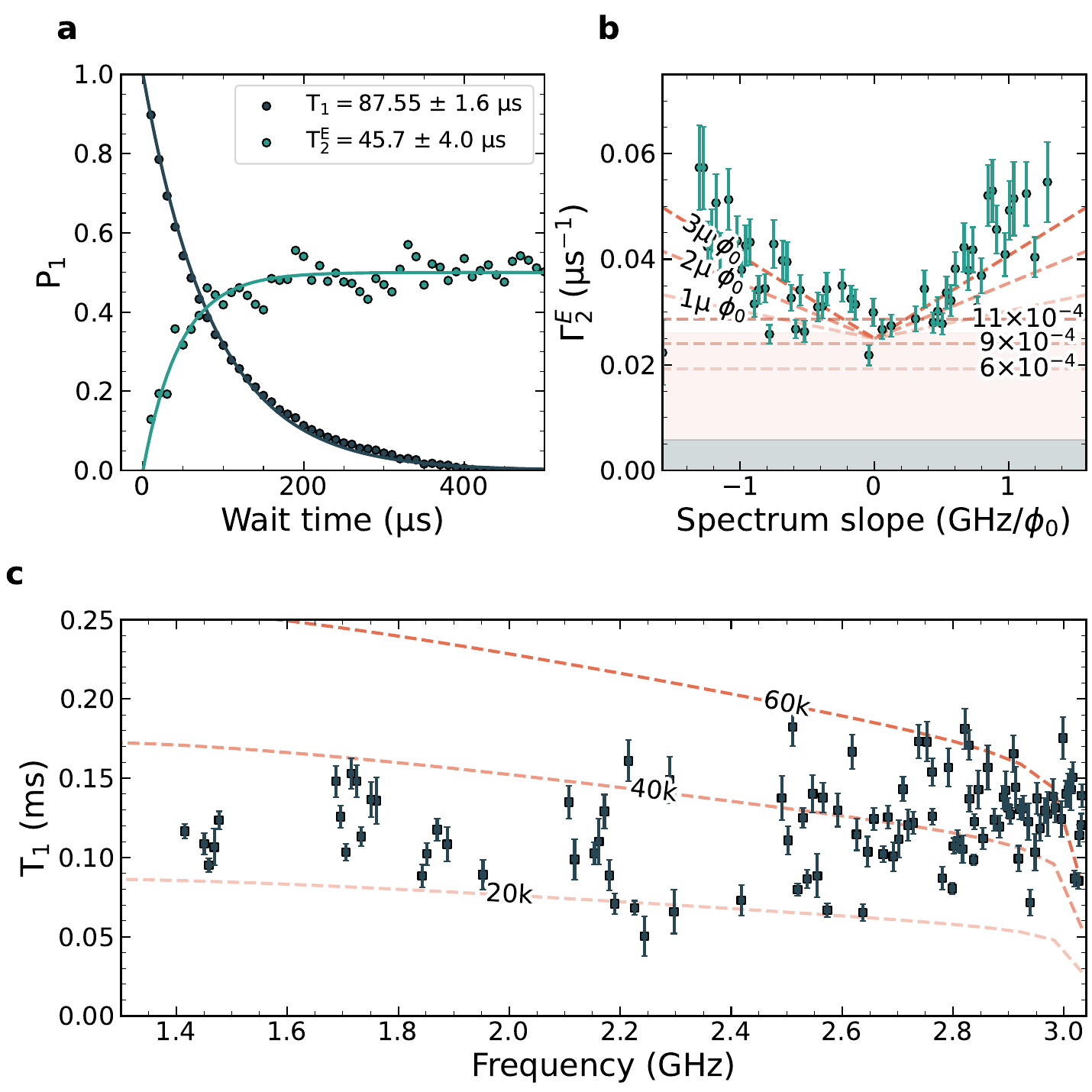}
\caption{\textit{Coherence of sample B}. \textbf{a)} Measured energy relaxation time $\text{T}_1$ and echo coherence time $\text{T}_{2}^E$. 
\textbf{b)} Measured echo decoherence rate $\Gamma_{2}^E$ as function of flux (light teal hexagonal markers), red dashed line represent the theory of dephasing due to 1/f flux noise for different noise amplitudes. Error bars are one standard deviation.
\textbf{c)} Measured energy relaxation time as function of external flux (dark teal square marks). Dashed lines represent the prediction due to dielectric losses for different dielectric quality factors.}
\label{coherencesampleB}
\end{figure}

Similarly to sample A, we performed a systematic characterization of sample B. The circuit spectrum was acquired using two-tone spectroscopy, and fitted to the fluxonium and dual Hamiltonian according to the procedure described in appendix \ref{spectro_fit} (see Fig.~\ref{Spec_sampleB}). The parameters obtained for sample B are $E_\text{C} = 1.62~\text{GHz}$, $E_\text{J} = 5.76~\text{GHz}$, and  $E_\text{L} = 162~\text{MHz}$ for the fluxonium Hamiltonian; and $E_\text{S,1} = 176~\text{MHz}$, $E_\text{S,2} = 178~\text{MHz}$ and $\text{E}_\text{L}^\star = 154 ~\text{MHz}$ for the dual Hamiltonian. The parameters of the two samples are quite close since both samples have been fabricated in the same batch. And the spectrum of both samples can be well explained by the dual model up to two excitations. The coherence at zero flux of sample B has also been characterized. Using the same methods as for sample A, we measured the depolarisation time $T_1 = 87.55 \pm 1.6 ~\text{µs}$, and the echo decoherence time $T_2^\text{E} = 45.7\pm 4~\text{µs}$ at integer flux (see Fig.~\ref{coherencesampleB}\textbf{a}).

To understand what limits the coherence time of this sample similar measurement have been repeated as a function of the external flux bias. $T_1$ also seems to be limited by capacitive loss (see Eq.~(\ref{capacitiveloss})). The dielectric quality factor $Q_\text{cap}$ is around 40 k (see Fig.~\ref{coherencesampleB}\textbf{c}). But given the large scatter in the dataset, together with a rather flat dependence with flux it is hard to conclude anything else that the capacitive losses and hence the depolarisation time of both samples have the same order of magnitude. 

We also measured the echo coherence time for various flux bias, as shown in Fig.~\ref{coherencesampleB}\textbf{b}. From this measurement we extract  $\sqrt{A_\phi} \sim 3$~\textmu$\phi_0$, which is close to the one reported for sample A and in agreement with the literature. As for sample A the flux noise does not seems to limit the coherence time of our qubit at zero flux. Accounting for photon shot noise we can reproduce the measured coherence time with $\bar{n}_\text{th} = 9
\times 10^{-4}$. This convert in an effective temperature $T_\mathrm{eff}$ of about $47$~mK which is close to the one measured for sample A.

\section{Dephasing caused by the second excited state}
\label{f-dephasing}

In this section, we study the impact of the second excited state on the coherence time of the qubit. As the second excited state is very close to the first excited state in energy, the transition $\lvert 1 \rangle \to \lvert 2\rangle$ can be thermally activated. To comprehend the impact of this hot transition on the coherence time, we write the Lindblad equation for the first three levels
\begin{equation}
    \frac{\partial \rho}{\partial t} = -i \left[ {H}, \rho \right] + \sum\limits_i D_i(\rho)
\end{equation}
Here, $\rho$ represents a qutrit density matrix, $H$ denotes the Hamiltonian, and $D_i(\rho)$ is the dissipator defined as:
\begin{equation}
    D_i(\rho) = L_i \rho L_i^\dag -\frac{1}{2} \{ L_i^\dag L_i , \rho \}
\end{equation}
With $L_j$ are a set of jump operators between the qutrit states.
To simplify the calculation,  we initially employ the rotating frame approximation, setting the $\left[ {H}, \rho \right] $ term to zero.  Furthermore, we consider an initial condition where the system is in the state $\lvert \psi_0 \rangle = (\lvert 0 \rangle+ \lvert 1 \rangle)/\sqrt{2}$. Our objective is to determine the duration for which the superposition is maintained while accounting for the effects of the jump operators:
\begin{align}
    & L_1 = \sqrt{\Gamma_{12}^\uparrow} \lvert 2\rangle \langle 1\rvert \\
    & L_2 = \sqrt{\Gamma_{12}^\downarrow} \lvert 1\rangle \langle 2\rvert
\end{align}
Here, $\Gamma_{12}^\uparrow$ and $\Gamma_{12}^\downarrow$ are the corresponding rates of excitation and relaxation between the states $\lvert 1\rangle$ and $\lvert 2 \rangle$. Since states $\lvert 0\rangle$ and $\lvert 2 \rangle$ share the same symmetry in the phase space, we neglect any jumps between them.
\begin{figure}
\centering
\includegraphics[width=0.5\columnwidth]{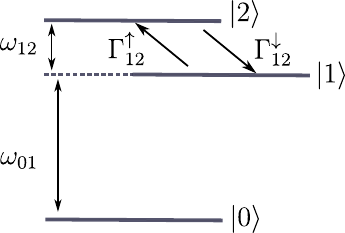}
\caption{\textit{Qutrit energy diagram}. Fluxonium at integer flux seen as a qutrit.}
\label{lindblad_fig}
\end{figure}
We parameterize the density matrix $\rho $ as follows:
\begin{align}
  \rho(t) = \begin{pmatrix}
    \rho_{00}(t) & \frac{x(t) - i y(t)}{2}& 0\\
    \frac{x(t) + i y(t)}{2} & \rho_{11}(t) & 0 \\
    0& 0 & \rho_{22}(t)
\end{pmatrix}
\end{align}
The jump operators yield the following dissipators
\begin{align}
    & D_1(\rho) = \Gamma_{12}^\uparrow \rho_{11}  \lvert 2\rangle \langle 2\rvert  - \frac{1}{2} \Gamma_{12}^\uparrow  (\lvert 1\rangle \langle 1\rvert \rho  + \rho \lvert 1\rangle \langle 1\rvert )\\
    & D_2(\rho) = \Gamma_{12}^\downarrow \rho_{22}  \lvert 1\rangle \langle 1\rvert  - \frac{1}{2} \Gamma_{12}^\downarrow  (\lvert 2\rangle \langle 2\rvert \rho  + \rho \lvert 2\rangle \langle 2\rvert )
\end{align}
Upon substituting these dissipators into the Lindblad equation, we obtain the following equations of motion:
\begin{align}
    & \frac{\partial \rho_{00}}{\partial t} = 0 \\
    & \frac{\partial \rho_{11}}{\partial t} = - \Gamma_{12}^\uparrow \rho_{11}  + \Gamma_{12}^\downarrow \rho_{22}\\
    & \frac{\partial \rho_{22}}{\partial t} = - \Gamma_{12}^\downarrow \rho_{22}  + \Gamma_{12}^\uparrow \rho_{11}\\
    &\frac{\partial x(t) }{\partial t} = -\frac{ \Gamma_{12}^\uparrow } {2} x(t)\\
    &\frac{\partial y(t) }{\partial t} = -\frac{ \Gamma_{12}^\uparrow } {2} y(t)
\end{align}
From the system's equations of motion, we deduce that only ${ \Gamma_{12}^\uparrow } /{2} $ contributes to the Ramsey decay rate on the qubit. Notably $\Gamma_{12}^\downarrow$ does not play a role in this context.

\bibliography{Bifluxon_biblio}

\end{document}